\documentclass[pra,twocolumn,showpacs,preprintnumbers,amsmath,amssymb,aps]{revtex4}
\usepackage{color}
\usepackage{bm, graphicx, amsmath}
\usepackage{hyperref}

\begin{document}
\title{Partial particle and wave information and weak duality games}
\author{Mark Hillery}
\affiliation{Department of Physics, Hunter College of the City University of New York, 695 Park Avenue, New York, NY 10065 USA \\ Physics Program, Graduate Center of the City University of New York, 365 Fifth Avenue, New York, NY 10016}

\begin{abstract}
Duality games are a way of looking at wave-particle duality.  In these games. Alice and Bob together are playing against the House.  The House specifies, at random, which of two sub-games Alice and Bob will play.  One game, Ways, requires that they obtain path information about a particle going through an $N$-path interferometer and the other, Phases, requires that they obtain phase information.  In general, because of wave-particle duality, Alice and Bob cannot always win the overall game.  However, if the required amount of path and phase information is not too great, for example specifying a set of paths or phases, one of which is the right one, then they can always win.  Here we study examples of duality games that can always be won, and develop a wave-particle duality relation expressed only in terms of mutual information to help analyze these games.
\end{abstract}

\maketitle

\section{Introduction}
Wave particle duality is a prominent feature of quantum mechanics.  Crudely put, it states that a quantum system can have wave-like properties, or particle like properties, but not both at the same time.  More refined statements express a tradeoff between wave and particle properties.  One of the earliest was a study by W.\ Wootters and W.\ Zurek of the proposal debated by Einstein and Bohr to use a movable screen with two slits to detect which path a photon took on its way to an observation screen \cite{wootters}.  The underlying model for most later treatments is a particle going, via multiple paths, through an interferometer.  Particle behavior is characterized by information about which path the particle took, and wave behavior is characterized by an interference measure.  In earlier studies, path information was determined by the probability distribution for the paths, and this was determined solely by the quantum state of the particle \cite{greenberger,jaeger}.  Path detectors were added to the model by Englert, and the overlap of the detector states determines the amount of path information one has \cite{englert}.    In the case of only two paths, the interference measure was taken to be the visibility of the interference pattern, but when the number of paths is increased, more complicated measures are called for \cite{durr,bimonte,englert2,jakob1,englert3}.  It was also found that entanglement between the particle and the path detectors can be incorporated into the duality relation, turning it into a ``triality'' relation \cite{jakob2}.  Most recently, coherence measures coming from the the resource theory of coherence \cite{baumgratz} have been used to formulate duality relations \cite{pati,bagan1,bagan2,bagan3,bagan4}.

Another way of treating the coherence measure is to have several sets of phases that can be applied to the different paths, and to try to determine which set of phases has been applied \cite{bagan3}.  In that case, the coherence measure becomes the probability with which we can successfully determine which set of phases has been applied, and the wave-particle duality relation becomes an upper bound on the sum of two probabilities, the probability of successfully determining the phase set and the probability of successfully determining the path.  The problem is simplified if the sets of phases comprise a representation of a cyclic group, and then each group element corresponds to a different set of phases. In this case, the coherence measure based on the probability of successfully discriminating the phases is equal to the $l_{1}$ coherence measure that comes from the resource theory of coherence \cite{baumgratz,bagan3}.  

Yet one more way of discussing duality relations is in terms of a game \cite{bagan2,coles1,coles2}.   The game consists of the House, who supervises the game, and two players, Alice and Bob.   There are two sub-games, one requiring path information and one requiring phase information, and each of these is played half the time.  For each round, the House determines which sub-game will be played, with each being equally likely, and Alice and Bob try to win it.  Their probability of winning the overall game is less than one, due to the incompatibility between phase and path information.

Here we will first look at a modified version of this game that requires Alice and Bob to acquire less path and phase information that does the original one.  This will allow them to win with a probability of one.  We will then examine how much phase and path information Alice and Bob can acquire and still always win the game.  One tool that will be useful in this endeavor is a wave-particle duality relation that is expressed purely in terms of mutual information.  

\section{Modified duality game}
Let us start by describing the original game.  The House prepares a state on $\mathcal{H}_{p}\otimes \mathcal{H}_{d}$ of the form ($U(g)\otimes I_{d})|\Psi\rangle$, where
\begin{equation}
\label{state}
|\Psi\rangle = \frac{1}{\sqrt{N}} \sum_{j=0}^{N-1} |j\rangle_{p}|\eta_{j}\rangle_{d} ,
\end{equation}
and gives the $p$ part to Alice and the $d$ part to Bob.  This corresponds to giving Alice access to the particle going through the interferometer and Bob access to the path detectors. Here $\{ |j\rangle_{p} \, |\, j=0,1,\ldots N-1\}$ is an orthonormal basis of $\mathcal{H}_{p}$, and these can be thought of as paths.  The states $\{ |\eta_{j}\rangle_{d} \, |\, j=0,1,\ldots N-1 \}$ are in $\mathcal{H}_{d}$ and represent states of the detectors coupled to the paths.  If the $|\eta_{j}\rangle_{d}$ are orthonormal, then we have perfect path information, because the detector states corresponding to different paths are perfectly distinguishable.  If they are not, then the detectors provide less information about the paths.  The operator $U(g)$ applies a set of phases to the states $\{ |j\rangle_{p} \, |\, j=0,1,\ldots N-1\}$, and we shall assume that it is an element of a representation of the group $\mathbb{Z}_{N}$, where $g\in \mathbb{Z}_{N}$.

As mentioned in the Introduction, the game consists of two parts.  If the House tells Alice and Bob to play Phases, Alice measures her system in order to determine $U(g)$, Bob does nothing, and Alice and Bob win if Alice is correct.  If the House tells them to play Ways, then Alice measures her system in the basis $\{ |j\rangle_{p} \, |\, j=0,1,\ldots N-1\}$ and Bob measures his, and they win if they get they same answer.  Phases and Ways are equally likely,, each occurs with a probability of $1/2$.  The probability that Alice and Bob win the overall game, $P_{win}$, is bounded by \cite{bagan2}.
\begin{equation}
P_{win} \leq \frac{1}{2} + \frac{1}{2\sqrt{N}} .
\end{equation}

We now want to modify the game, by requiring less information from Alice and Bob, so that they can always win.  We will do this via an example.  Consider an interferometer with three paths, and let $U(g)$ be a representation of $\mathbb{Z}_{3} = \{ e,a,a^{2} \}$, where $a^{3}=e$, with $U(e) = I_{p}$, $U(a)|j\rangle_{p} = \exp (2\pi ij/3) | j\rangle_{p}$, and $U(a^{2}) = U^{2}(a)$, where $I_{p}$ is the identity operator on the path space.  These three operators constitute a three dimensional representation of the group $\mathbb{Z}_{3}$, with each irreducible representation appearing once, and the states $|j\rangle_{p}$ being the one-dimensional invariant subspaces of the irreducible representations.  The basic idea is that instead of trying to find which operator, $U(g)$ was applied or which state Alice found when she measured her path states, one wants to find sets that contain the desired operation or state. In particular, if they play Phases, Alice wants to identify a set of two operations one of which is the one that was applied.  If they play Ways, Bob wants to find a set containing two possible path states, one of which is the one Alice found.  In each case two possibilities remain, so while Alice and Bob have gained some information about a path or a set of phases, they have gained less than if they tried to determine exactly the path or the set of phases.  

Note that each of these measurements can be accomplished by eliminating one out of three possibilities.  Elimination measurements have been an area of recent attention \cite{PBR,Caves,Perry,Bando,Heinosaari,Crickmore,andersson}.  Suppose they are playing Ways, and Bob's measurement is guaranteed to give a result different from the one Alice got.  For example, if Alice got $0$, Bob could get either $|\eta_{1}\rangle_{d}$ or $|\eta_{2}\rangle_{d}$.  He then names the two results he did not get, for example, if he got $|\eta_{1}\rangle_{d}$ he would specify $\{ 0,2\}$, and that set will contain Alice's result.  Similarly, if they play Phases and Alice's measurement gives an operation $U(g)$ that was not applied, then she just names the other two, and one of these is guaranteed to be correct.  These are both examples of exclusion measurements, measurements that excludes possibilities.  
  
Now let's show that the modified game can be won all the time.  We choose the detector to be a qubit and for the detector states, we choose the trine states
\begin{eqnarray}
|\eta_{0}\rangle_{d} & = & |0\rangle \nonumber \\
|\eta_{1}\rangle_{d} & = & -\frac{1}{2} |0\rangle + \frac{\sqrt{3}}{2} |1\rangle \nonumber \\
|\eta_{2}\rangle_{d} & = & -\frac{1}{2} |0\rangle - \frac{\sqrt{3}}{2} |1\rangle .
\end{eqnarray}
Bob's POVM will be based on the anti-trine states 
\begin{eqnarray}
|\bar{\eta}_{0}\rangle_{d} & = & |1\rangle \nonumber \\
|\bar{\eta}_{1}\rangle_{d} & = & -\frac{\sqrt{3}}{2} |0\rangle -\frac{1}{2} |1\rangle \nonumber \\
|\bar{\eta}_{2}\rangle_{d} & = & \frac{\sqrt{3}}{2} |0\rangle -\frac{1}{2} |1\rangle .
\end{eqnarray}
Note that $\,_{d}\langle\eta_{j}|\bar{\eta}_{j}\rangle_{d} = 0$.  His measurement operators are given by $A_{j}=\sqrt{2/3} |\bar{\eta}_{j}\rangle_{d}\langle\bar{\eta}_{j}|$ with POVM elements $\Pi_{j}=A_{j}^{\dagger}A_{j}$.  This will guarantee that if Alice and Bob play Ways, Bob will get a result different from Alice's, and by naming the two states he did not get, one of which will be the state Alice sent, he and Alice will win Ways.  For example, suppose Alice got $1$ as the result of her measurement, corresponding to the state $|1\rangle_{p}$.  Bob's result will be either $0$ or $2$, corresponding to the states $|\bar{\eta}_{0}\rangle_{d}$ or $|\bar{\eta}_{2}\rangle_{p}$, respectively.  If he got $0$, he names the pair $(1,2)$ and if he gets $2$, he names the pair $(0,1)$, and both of these pairs contain the result of Alice's measurement.

What happens if they play Phases?  Since only Alice will be making the measurement, we trace out the detector states to get the path density matrix,
\begin{eqnarray}
\rho_{p} & = & \frac{1}{3} \sum_{j,k=0}^{2} \,_{d}\langle \eta_{k}|\eta_{j}\rangle_{d} |j\rangle_{p}\langle k| \nonumber \\
& = & \frac{1}{2} ( |u_{1}\rangle_{p}\langle u_{1}| + |u_{2}\rangle_{p}\langle u_{2}| ) .
\end{eqnarray}
Here
\begin{eqnarray}
|u_{0}\rangle_{p} & = & \frac{1}{\sqrt{3}} \sum_{j=0}^{2} |j\rangle_{p}  \nonumber \\
|u_{1}\rangle_{p} & = & \frac{1}{\sqrt{3}} \sum_{j=0}^{2} e^{2\pi i j/3} |j\rangle_{p}  \nonumber \\
|u_{2}\rangle_{p} & = & \frac{1}{\sqrt{3}} \sum_{j=0}^{2} e^{-2\pi ij/3} |j\rangle_{p}  .
\end{eqnarray}
Note that
\begin{eqnarray}
U(a) \rho_{p} U^{-1}(a) & = & \frac{1}{2} ( |u_{0}\rangle_{p}\langle u_{0}| + |u_{2}\rangle_{p}\langle u_{2}| )  \nonumber  \\
U(a^{2})\rho_{p} U^{-1}(a^{2}) & = & \frac{1}{2} ( |u_{0}\rangle_{p}\langle u_{0}| + |u_{1}\rangle_{p}\langle u_{1}| )  .
\end{eqnarray}
Alice can measure in the basis $\{ |u_{j}\rangle_{p}\, | j=0,1,2 \}$.  If Alice gets $|u_{0}\rangle_{p}$, then she knows that $|\Psi\rangle$ was not sent, if she gets $|u_{1}\rangle_{p}$, she knows that $U(a)\otimes I_{d} |\Psi\rangle$ was not sent, and if she gets $|u_{2}\rangle_{p}$, then she knows that $U(a^{2})\otimes I_{d} |\Psi\rangle$ was not sent.  So, she can name the two states that were not eliminated by her measurement result, one of which will be the state that was sent, thereby winning Phases.  

In conclusion, if the House specifies either Ways or Phases, with the modified rules Alice and Bob can always win the game.  This is because the new rules do not require them to acquire too much path and phase information.  In the next section we will find a relation that places on upper bound on the amount of path and phase information one can have, and allows us to place limits on the kinds of games that can be won all the time.

\section{Duality relation based on mutual information}

We would like to examine other wave-particle duality games and see which ones can be won all the time.  The expectation is that this will not be possible if too much wave and particle information is required, but how much is too much?  A useful tool in studying this question is a wave-particle duality relation that is expressed purely in terms of mutual information.  

The basic setup is the same as in the previous section, but now with the initial state
\begin{equation}
|\Psi_{g}\rangle = \sum_{j=0}^{N-1} \sqrt{p_{j}} \, U(g)|j\rangle_{p} |\eta_{j}\rangle_{d}  ,
\end{equation}
for some element of a group representation, $U(g)$ and $\sum_{j=0}^{N-1}p_{j}=1$.  Let $X_{1}$ be the random variable corresponding to the state that was sent, and $X_{2}$ be the random variable corresponding to the result of the measurement Alice makes in order to gain information about that state.  Similarly, $Y_{1}$ is the random variable corresponding to the measurement Alice makes to determine the path, a von Neumann measurement with projections $|j\rangle_{p}\langle j|$, and $Y_{2}$ is the random variable corresponding to the result of Bob's measurement on the detector states. 

Let us now be more specific about the group.  We will assume it is $\mathbb{Z}_{N}$, and that $U(g)$ is an $N$-dimensional representation, which is a direct sum of the irreducible representations of $\mathbb{Z}_{N}$ with each irreducible representation appearing once.  The generator of the group is $a$, and we have that $U(a^{k})|j\rangle_{p} = \exp (2\pi i jk/N) |j\rangle_{p}$, where $k\in \{ 0, 1,\ldots N-1\}$.  We will assume that each group element is equally likely.  The density matrices that Alice must try to distinguish are $\rho_{p}^{(k)} = U(a^{k}) \rho_{p}^{(0)} U^{-1}(a^{k})$, where
\begin{equation}
\rho_{p}^{(0)} = {\rm Tr}_{d}(|\Psi\rangle\langle\Psi |) = \sum_{j,k=0}^{N-1} \sqrt{p_{j}p_{k}} \,_{d}\langle \eta_{k}|\eta_{j}\rangle_{d} |j\rangle_{p}\langle k|   .
\end{equation}
Applying the Holevo bound to Alice's measurement, we have \cite{nielsen}
\begin{eqnarray} 
I(X_{1}:X_{2}) & \leq & S\left(\frac{1}{N}\sum_{k=0}^{N-1} \rho_{p}^{(k)} \right) -\frac{1}{N} \sum_{k=0}^{N-1} S(\rho_{p}^{(k)})   \nonumber \\
 & \leq & H(\{ p_{j}\}) - S(\rho_{p}^{(0)}) .
 \end{eqnarray}
Here we have used the fact that the first term on the right-hand side is equal to $S(\sum_{j=0}^{N-1}p_{j}|j\rangle_{p}\langle j|) = H(\{ p_{j}\})$, the Shannon entropy of the probability distribution $\{ p_{j}\, |\, j=0,1,\ldots N-1 \}$, and the entropies of the density matrices $\rho_{p}^{(k)}$ are all the same, because they are related by unitary operators.  Note that $H(\{ p_{j}\}) - S(\rho_{p}^{(0)})$ is just the relative-entropy measure of coherence of $\rho_{p}^{(0)}$.  Next we turn to Bob's measurement.  Again applying the Holevo bound we have
 \begin{equation}
 I(Y_{1}:Y_{2}) \leq S\left( \sum_{j=0}^{N-1} p_{j}|\eta_{j}\rangle_{p}\langle\eta_{j}|\right) .
 \end{equation}
If we now add these two inequalities and use the fact that the entropies of the reduced density matrices of a pure bipartite state are equal, we find that 
\begin{equation}
\label{mut-inf-dual}
I(X_{1}:X_{2}) + I(Y_{1}:Y_{2}) \leq H(\{ p_{j}\})  .
\end{equation}
Note that we have specified neither Alice's measurement to gain information about which state was sent nor Bob's measurement.  Information based duality relations have appeared in \cite{bagan1,bagan4,angelo}.

As an example, let's see if this relation is consistent with the situation described in the previous section.  Recall that the mutual information between two random variables, $X$ and $Y$ is given by
\begin{equation}
I(X:Y) = \sum_{x,y} p(x,y)\log\left[ \frac{p(x,y)}{p_{X}(x)p_{Y}(y)}\right] ,
\end{equation}
where $p(x,y)$ is the joint distribution of $X$ and $Y$, $p_{X}(x)$ is the distribution for $X$, $p_{Y}(y)$ is the distribution for $Y$, and logarithm is base $2$.  We will start by computing the mutual information of a measurement that eliminates one of three possibilities.  We have two random variables, $Z_{1}$ and $Z_{2}$, that both take values in the set $\{ 0,1,2\}$.  We will assume that $p(z_{1}) = 1/3$ and $p(z_{2}|z_{1}) = (1/2)(1-\delta_{z_{1}z_{2}})$.  This gives us that $p(z_{2})=1/3$.  The mutual information between $Z_{1}$ and $Z_{2}$ is, $I(Z_{1}:Z_{2}) = \log 3 -1$.  In the example we considered in the previous section, both measurements were of this type, so $I(X_{1}:X_{2}) + I(Y_{1}:Y_{2}) = 2(\log 3 -1) = \log(9/4)$ and $H(\{ p_{j}\}) = \log 3$. so the inequality is satisfied.  Note that if one of pairs of random variables is perfectly correlated, then their mutual information would be $\log 3$, so the mutual information of the other pair must be zero.  

\section{Four paths}
We will look at two examples of four paths.  In both we will perform a path measurement that reduces the number of paths from four to two.  In one case there will be $6$ possible measurement outcomes, each corresponding to one of the possible pairs of paths.  In the second there will be only two possible outcomes, each corresponding to one of two pairs.

Let us begin by finding the mutual information between two random variables, $X$ and $Y$, where $X\in \{ 0,1,2,3 \}$ labels the paths, and $Y\in \{ S_{k}\, |\, k=1,2, \ldots 6\}$ labels the possible pairs ($S_{j}$ is a pair of paths).  We will assume that each path is equally likely, so $p_{X}(x)=1/4$.  The conditional probability $p(y|x)$ is
\begin{equation}
\label{6pairs}
p(y|x) = \left\{ \begin{array}{cc} 1/3 & \hspace{5mm} x\in S_{y} \\ 0 & \hspace{5mm} {\rm otherwise} \end{array} \right.
\end{equation}
Note that each $x$ can be in three possible pairs, and we are assuming that each pair is equally likely.  
From this we find that $p_{Y}(y) = 1/6$, and $I(X:Y)= 1$.  If we assume that Alice's phase measurement does the same thing, eliminates two out of four possibilities, then the two mutual informations in Eq.\ (\ref{mut-inf-dual}) are both equal to one and $H(\{ p_{j}\}) = 2$, because $p_{j}=1/4$ for $j=0,1,2,3$.  Therefore, this set of path and phase measurements does not violate Eq.\ (\ref{mut-inf-dual}).  If Alice's phase measurement is weaker, eliminating only one possible set of phases, then that is also allowed, and this is what we shall find: a path measurement that eliminates two possible paths and a phase measurement that eliminates one possible phase set.

Our detector space, $\mathcal{H}_{d}$, is now three dimensional, with an orthonormal basis $\{ |0\rangle _{d}, |1\rangle_{d} , |2\rangle_{d} \}$.  The detector states are
\begin{eqnarray}
|\eta_{0}\rangle_{d} & = & |2\rangle_{d}  \nonumber \\
|\eta_{1}\rangle_{d} & = & \frac{2\sqrt{2}}{3} |0\rangle_{d} - \frac{1}{3} |2\rangle_{d}  \nonumber \\
|\eta_{2}\rangle_{d} & = & -\frac{\sqrt{2}}{3} |0\rangle_{d} + \sqrt{\frac{2}{3}} |1\rangle_{d} -\frac{1}{3} |2\rangle_{d}  \nonumber \\
|\eta_{3}\rangle_{d} & = & -\frac{\sqrt{2}}{3} |0\rangle_{d} - \sqrt{\frac{2}{3}} |1\rangle_{d} -\frac{1}{3} |2\rangle_{d} .
\end{eqnarray}
Note that $\,_{d}\langle\eta_{j}|\eta_{k}\rangle_{d} = -1/3$ for $j\neq k$.  In order to find a POVM whose elements correspond to detecting pairs of these states, we will find vectors that are orthogonal to different pairs.  These states are, where the subscript indicates the pair to which the state is orthogonal, 
\begin{eqnarray}
|\xi_{01}^{\perp}\rangle_{d} & = & |1\rangle_{d}  \nonumber \\
|\xi_{02}^{\perp}\rangle_{d} & = & -\frac{\sqrt{3}}{2} |0\rangle_{d} - \frac{1}{2} |1\rangle_{d} \nonumber \\
|\xi_{03}^{\perp}\rangle_{d} & = & \frac{\sqrt{3}}{2} |0\rangle_{d} - \frac{1}{2} |1\rangle_{d} \nonumber \\
|\xi_{12}^{\perp}\rangle_{d} & = & \frac{\sqrt{3}}{2}\left( \frac{1}{3} |0\rangle_{d} + \frac{1}{\sqrt{3}}|1\rangle_{d} + \frac{2\sqrt{2}}{3} |2\rangle_{d} \right) \nonumber \\
|\xi_{13}^{\perp}\rangle_{d} & = & \frac{\sqrt{3}}{2}\left( \frac{1}{3} |0\rangle_{d} - \frac{1}{\sqrt{3}}|1\rangle_{d} + \frac{2\sqrt{2}}{3} |2\rangle_{d} \right) \nonumber \\
|\xi_{23}^{\perp}\rangle_{d} & = & \frac{\sqrt{3}}{2} \left( \frac{2}{3} |0\rangle_{d} - \frac{2\sqrt{2}}{3} |2\rangle_{d} \right) .
\end{eqnarray}
We can now define a six element POVM that detects which pair a path state is in by
\begin{equation}
\Pi_{jk} = \frac{1}{2} |\xi^{\perp}_{(\overline{jk})}\rangle_{d} \langle\xi^{\perp}_{(\overline{jk})}| ,
\end{equation}
where $\overline{jk}$ denotes the pair that has no elements in common with $jk$.  Note that if the detector corresponding to the pair $jk$ clicks, that means that the path-detector state was not a member of the pair $\overline{jk}$, so it must have been a member of the pair $jk$.  We find that
\begin{equation}
\,_{d}\langle \eta_{j}|\Pi_{jk}|\eta_{j}\rangle_{d} = \,_{d}\langle \eta_{k}|\Pi_{jk}|\eta_{k}\rangle_{d} = \frac{1}{3} .
\end{equation}
This is consistent with the measurement specified in Eq.\ (\ref{6pairs}).  Using this measurement, Bob can find a pair of paths, one of which is the same as the one Alice found.

We now need to look at the phases.  Alice and Bob share the state in Eq.\ (\ref{state}) with $N=4$.  The House applies one of the operators $U(a^{k})$ for $k=0,1,\ldots 3$, to Alice's part of the state, where $U$ is a representation of $\mathbb{Z}_{4} = \{ a^{k}\, |\, k=0,1,\ldots 3\}$.  In particular, $U(a^{k}) |j\rangle_{p}= \exp (i\pi jk/2) |j\rangle_{p}$.  In order to determine what measurement Alice should make to determine which operator,, $U(a^{k})$, has been applied to her state, we first examine her reduced density matrix for $k=0$,
\begin{equation}
\rho_{p}^{(0)} = \frac{1}{4} I_{p} - \frac{1}{12} \sum_{j\neq k} |j\rangle_{p}\langle k| .
\end{equation}
Diagonalizing this density matrix we find
\begin{equation}
\rho_{p}^{(0)} = \frac{1}{3} \sum_{j=1}^{3} |v_{j}\rangle_{p}\langle v_{j}| ,
\end{equation}
where
\begin{equation}
|v_{j}\rangle_{p} = \frac{1}{2} \sum_{k=0}^{3} e^{i\pi jk/2} |k\rangle_{p} ,
\end{equation}
where $j=0,1,\ldots 3$.  Note that $U(a^{k}) |v_{j}\rangle_{p} = |v_{j+k}\rangle_{p}$, where the addition is modulo $4$.  The situation here is very similar to what we saw with the trine states.  Each of the density matrices, $\rho_{p}^{(k)} = U(a^{k}) \rho_{p}^{(0)} U^{-1}(a^{k})$ has one of the vectors $|v_{j}\rangle_{p}$ missing, in particular, $\rho_{p}^{(k)}$ has $|v_{k}\rangle_{p}$ missing.  Therefore, if Alice measures in the basis $\{ |v_{j}\rangle_{p}\, |\, j=0,1,\ldots 3\}$, she is able to eliminate one of the possible operators, $U(a^{k})$.  

This implies that Alice and Bob can always win the following game.  If they play Ways, Bob must name a pair of paths, one of which is the same as the one Alice found.  If they play Phases, Alice must name three possible phase sets, one of which is the right one.  Because Alice cannot eliminate a pair of operators with the scheme, it will not saturate the inequality in Eq.\ (\ref{mut-inf-dual}).

We can saturate the inequality with a simpler scheme.  Consider only two path pairs, $\{ 1,3\}$ and $\{ 0,2\}$.  Our detector will be a qubit, and the state shared by Alice and Bob will be $U(a^{k}) \otimes I_{d} |\Psi\rangle$, where
\begin{equation}
|\Psi\rangle = \frac{1}{2} (|1\rangle_{p} + |3\rangle_{p})|0\rangle_{d} + \frac{1}{2} (|0\rangle_{p} + |2\rangle_{p}) |1\rangle_{d}  .
\end{equation}
Because the detector states are orthogonal, Bob can determine which pair of states will contain the result of Alice's measurement.  Using the fact that
\begin{eqnarray}
|1\rangle_{p} + |3\rangle_{p} & = & |v_{0}\rangle_{p} - |v_{2}\rangle_{p}   \nonumber \\
 |0\rangle_{p} + |2\rangle_{p} & = & |v_{0}\rangle_{p} + |v_{2}\rangle_{p}  ,
\end{eqnarray}
we find that for $k=0$ Alice's density matrix is
\begin{equation}
\rho_{p}^{(0)}  = \frac{1}{2} (|v_{0}\rangle_{p}\langle v_{0}| + |v_{2}\rangle_{p}\langle v_{2}|) .
\end{equation}
From this, we have that $\rho_{p}^{(0)} = \rho_{p}^{(2)}$ and $\rho_{p}^{(1)} = \rho_{p}^{(3)}$.  Therefore, if Alice measures in the $\{ |v_{j}\rangle_{p}\, |\, j=0,1,\ldots 3\}$ basis, she is able to identify a pair one of whose members is the unitary operator that was applied.  In this case the mutual information for both the phase and path measurements is $1$, $H(\{ p_{j}\}) =2$, so our mutual information duality relation is saturated.

\section{N Paths}
We now want to see what kind of restrictions the duality relation places on the kinds of path and phase measurements that are possible.  Alice and Bob are now given the state $U(a^{k})\otimes I_{d} |\Psi\rangle$, where $|\Psi\rangle$ is the state in Eq.\ (\ref{state}), the group is $\mathbb{Z}_{N} =\{  a^{k}\, |\, k=0,1, \ldots N-1\}$, and $U(a^{k})|j\rangle_{p} = \exp (2\pi i jk/N) |j\rangle_{p}$.  Suppose that $N=mn$, where $n$ and $m$ are integers.  We want to split both the paths and operators, $U(a^{k})$, into $n$ sets of size $m$, and we want our measurements to tell us which set the path or the operator is in.  

Both the phase and path measurements are then of the following type.  There two random variables, $X\in \{ 0,1,\ldots N-1\}$ and $Y\in \{ 1,2,\ldots n\}$, where $Y$ labels the sets of size $m$, $S_{y}$, and these sets have no common elements.  We have that $p_{X}(x)=1/N$ and 
\begin{equation}
p(y|x) = \left\{ \begin{array}{cc} 1 & \hspace{5mm} x\in S_{y} \\ 0 & \hspace{5mm} {\rm otherwise} \end{array} \right.
\end{equation}
This implies that $p_{Y}(y)= m/N$, and the mutual information is $I(X:Y)= \log n$.  Now since both the phase and path measurements are of this type, the duality relation gives us that
\begin{equation}
2\log n \leq \log N ,
\end{equation}
which implies that $n\leq \sqrt{N}$.  This implies that the number of sets cannot be too large.  For example, if $N=6$, three sets of two elements each will not work, but two sets of three elements each may.

\section{Conclusion}
We have used wave-particle duality games to investigate how much path and phase information one can have at the same time.  The House decides whether Alice and Bob have to provide path information or phase information, and if the amount of information they have to provided about the path or phase is not too great, they can always win the game.  The measurements that Alice and Bob make identify sets of outcomes that contain the correct outcome, and the amount of information they obtain depends on the size of the sets.  We first looked at an example with three paths and three sets of phases, where Alice and Bob are able to eliminate one of the possibilities in each case.  We then went on to look at more complicated examples. In addition, we found a wave-particle duality relation expressed solely in terms of mutual information, and we were able to use this relation to find situations in which Alice and Bob cannot always win the duality game.

.

 \end{document}